\documentclass[aps,pra,reprint,superscriptaddress,longbibliography,nofootinbib]{revtex4-2}
\usepackage[T1]{fontenc}
\usepackage{lmodern,amsmath,amssymb}
\usepackage[colorlinks=true,linkcolor=blue,citecolor=blue,urlcolor=blue,breaklinks=true]{hyperref}
\usepackage{orcidlink}

\begin{document}

\title{Quantum nonclassicality as a marginal problem}

\author{Karl Svozil\,\orcidlink{0000-0001-6554-2802}}
\email{karl.svozil@tuwien.ac.at}
\affiliation{Institute for Theoretical Physics, TU Wien, Wiedner Hauptstra{\ss}e 8-10/136, A-1040 Vienna, Austria}

\date{\today}

\begin{abstract}
Classical explanations of quantum statistics---local, noncontextual, or causal---fail for one and the same reason: the probability distributions observed in different contexts are not the marginals of a single joint distribution.
Every single context admits a classical model.
This is Boole's problem of the ``conditions of possible experience.''
Garg and Mermin treated it as a linear problem and showed, via Farkas's lemma, that its solvability is equivalent to a finite list of Bell-type inequalities, one for each extreme ray of a convex polyhedral cone; Pitowsky gave it the form of correlation polytopes.
We sketch this common structure and discuss two examples:
Aerts's elastic-band model, which reproduces the quantum cosine in one context but obeys the classical bound when contexts are combined;
and nonclassicality in causal data fusion, where the contexts are supplied by interventions on a causal structure.
\end{abstract}

\maketitle

\section{Introduction}

Statements that quantum correlations ``cannot be explained classically'' are claims about several contexts at once.
In any single context---a single pair of settings in a Bell experiment, a single maximal set of comeasurable observables---the observed distribution is trivially reproduced by a classical random variable.
Unperformed experiments have no results~\cite{peres222}, and the classical difficulty only arises when one asks for values, or a probability distribution, that cover all contexts simultaneously.
The purpose of this note is to recall that this question has a precise and old formulation---the marginal problem---and that nonclassicality of quantum data, whatever the source of its contexts, is a failure of it.

\section{The marginal problem}
\label{sec:marginal}

Let $X_1,\ldots ,X_n$ be observables and let $\mathcal{C}=\{C_1,\ldots ,C_m\}$ be a family of contexts, that is, subsets of observables measured together.
Suppose that for each context $C$ a distribution $p_C$ of the values of the observables in $C$ is given, and that these distributions agree on common observables (no-signalling, or no-disturbance).
The marginal problem asks whether there is a single joint distribution $P(x_1,\ldots ,x_n)$ whose marginals are the $p_C$.

Boole formulated this as the question of the ``conditions of possible experience'' that frequencies of logically related events must obey~\cite{Boole-62,pitowsky-89a,Pit-94}; Froissart derived Bell-type inequalities constructively as faces of the convex hull of the classical, deterministic correlations~\cite{froissart-81}.
Whether consistent marginals always extend depends on how the contexts intertwine: Vorob'ev showed that they do, for all consistent data, if and only if the family of contexts is regular, essentially acyclic~\cite{Vorobev-1962}.
Nonclassicality therefore needs cycles of contexts.
The four contexts $\{A,B\}$, $\{A,B'\}$, $\{A',B\}$, $\{A',B'\}$ of the CHSH scenario form a closed cycle; any three of them form a chain and always admit a joint distribution.
In the two-setting (CHSH) scenario Fine showed that a local hidden-variable model exists if and only if such a joint distribution exists, and that the Clauser--Horne inequalities are then necessary and sufficient~\cite{Fine-82}. Garg and Mermin rederived this result with the general method of Sec.~\ref{sec:gm}, noted a gap in Fine's explicit construction, and pointed out that with more settings the two-setting conditions, imposed on every quadruple of variables, need not suffice~\cite{Garg1984}.
For the Kochen--Specker theorem the contexts are maximal sets of comeasurable observables intertwined through common elements~\cite{specker-60,kochen1}, and Abramsky and Brandenburger cast nonlocality and contextuality uniformly as locally consistent data without a global section~\cite{abramsky-brandenburger-2011}.

\section{Garg, Mermin, and Farkas's lemma}
\label{sec:gm}

Garg and Mermin gave this question its general linear form~\cite{Garg1984,Garg-Mermin-prl-84}.
Given pair distributions for some pairs of variables, they asked whether a specified set of higher-order distributions---for instance, only some of the third-order ones---can be constructed that is compatible with the given pairs, consistent on common variables, and non-negative.
A single joint distribution of all variables is the special case in which this set has one member.
They emphasized that the problem underlies all studies of local hidden-variable models built on the inequalities of Bell or of Clauser and Horne, and they formulated it without reference to quantum theory.

Two of their observations are central here.
First, compatibility and consistency alone can always be satisfied by a signed function; only non-negativity gives the problem its content.
A joint quasi-distribution of all variables always exists, and nonclassicality means that its negativity cannot be removed.
Second, even when non-negative functions exist, they need not have any meaning as distributions; what is informative is the case in which they cannot exist.

Write the compatibility and consistency conditions as linear equations $\sum_\mu v_{\mu\alpha}P_\mu = p_\alpha$ for the unknown entries $P_\mu$ of the higher-order distributions, where $v_{\mu\alpha}\in\{0,\pm 1\}$ and $p_\alpha$ is a given pair probability, or zero for a consistency condition.
By Farkas's lemma~\cite{Farkas1902}, non-negative $P_\mu$ exist if and only if
\begin{equation}
\begin{aligned}
&\textstyle\sum_\alpha c_\alpha p_\alpha \ge 0
\quad\text{for every } c \text{ with} \\
&\textstyle\sum_\alpha c_\alpha v_{\mu\alpha}\ge 0 \ \text{for all } \mu .
\end{aligned}
\label{eq:Farkas}
\end{equation}
The admissible $c$ form a convex polyhedral cone, so it suffices to check the finitely many inequalities $c^{(q)}\cdot p\ge 0$ for its extreme rays $c^{(q)}$; rays without negative components are satisfied automatically.
The necessary and sufficient conditions for classicality are thus a finite list of linear inequalities on the observed probabilities---Bell-type inequalities---and finding them amounts to finding the extreme rays of a cone, a task Garg and Mermin set up for exact computer solution and applied to the spin-$s$ version of the Einstein--Podolsky--Rosen--Bohm experiment~\cite{Garg1984}.
Pitowsky gave the dual, geometric picture: the classical models make up a correlation polytope, whose facets are Boole's conditions of possible experience~\cite{pitowsky-86,pitowsky,Pit-91,Pit-94}.
Complete facet sets for specific configurations have since been obtained by convex-hull computations~\cite{2000-poly}.

\section{Two examples}

\subsection{One context versus many: the elastic band}
\label{sec:band}

Classical models that reproduce quantum statistics in a single context show the same point from the other side.
In Aerts's elastic-band models~\cite{aerts:82,aerts-69,Aerts-91,Aerts_2014,Aerts_2016} the shared state is an elastic along a direction $J$ that breaks at a point $x$ distributed uniformly in $[-1,1]$.
A measurement along a direction at angle $\alpha$ to $J$ yields
\begin{equation}
A(\alpha;x)=\operatorname{sgn}(\cos\alpha-x),
\quad
P(A=+1)=\cos^2\frac{\alpha}{2},
\end{equation}
and two such observables read off the same share have
\begin{equation}
E(\alpha,\beta)=\langle A(\alpha;x)A(\beta;x)\rangle = 1-\vert\cos\alpha-\cos\beta\vert .
\end{equation}
For $\alpha=0$, that is, for one observable aligned with the share, this is the quantum form $E(0,\theta)=\cos\theta$~\cite{svozil-2022-epr}.

For a fixed share, however, all four outcomes $A(\alpha;x)$, $A(\alpha';x)$, $B(\beta;x)$, $B(\beta';x)$ are functions of the same $x$.
The model is a joint distribution over all contexts, every inequality of the form~\eqref{eq:Farkas} holds, and the CHSH expression obeys $\vert S\vert\le 2$: the cosine of the aligned context is compensated by the non-cosine correlations of the others~\cite{svozil-2022-epr}, as Peres had found for a classical exploding-bomb model~\cite{peres222}.
A uniform cosine in every context, and with it a violation of the CHSH inequality up to the Tsirelson bound and beyond~\cite{svozil-krenn,svozil-2022-epr}, requires re-aligning the share with whatever setting is chosen.
The hidden state then depends on the context, and different contexts are described by different distributions rather than by marginals of one.
The price is information about the context: one bit suffices~\cite{svozil-2022-epr}, compare the communication cost of simulating Bell correlations~\cite{toner-bacon-03,svozil-2004-brainteaser}, and an elastic-band device even realizes a Popescu--Rohrlich box~\cite{AertsSven,cerf-gisin-massar-pop-04,pop-rohr}.
Classical predictions for intertwined contexts are, moreover, contingent on the particular classical model chosen~\cite{svozil-2020-c}.

\subsection{Contexts from interventions: causal data fusion}

In causal inference a data table may be collected passively or under an intervention that sets a variable to a chosen value~\cite{pearl-causality}.
Lauand, Bekele, and Wolfe~\cite{Lauand-2024-datafusion,Lauand-2024-confounders} ask whether passive and interventional tables on a given causal structure admit a single classical causal model.
They find quantum examples in which each table separately has a classical model but the family has none, and present this as a new form of nonclassicality.
For binary interventions their counterfactual-graph construction turns the tables into one distribution over a graph with setting copies of the intervened variables: the passive table supplies the contexts in which the setting copy equals the value actually taken, the interventional tables the others.
Joint compatibility is then a marginal problem of the kind of Sec.~\ref{sec:marginal}.
Their facet inequalities, obtained by enumerating deterministic strategies and eliminating variables, are inequalities $c\cdot p\ge 0$ of the kind in Eq.~\eqref{eq:Farkas}.
Their cases in which every pair of tables is jointly classical but the triple is not---the triple violating a relabeled tripartite Bell inequality of \'Sliwa's list~\cite{sliwa-2003}---are instances of the pairwise-consistent but globally inconsistent families characterized by Vorob'ev.

Three features are nevertheless specific to interventions.
First, the passive and interventional regimes overlap: an interventional table also contains the runs in which the intervened variable would have taken the imposed value anyway, so the remaining contexts are obtained by subtracting the passive table rather than read off directly.
Second, when an intervened variable takes more than two values, the tables no longer determine the counterfactual distribution; the problem remains a marginal problem, now with fewer prescribed marginals, and its facets need not coincide with known Bell inequalities~\cite{Lauand-2024-datafusion}.
Third, several independent latent sources impose nonlinear independence constraints, so the classical set is no longer a polytope and Eq.~\eqref{eq:Farkas} does not apply directly; there Lauand, Bekele, and Wolfe use the inflation technique and quadratic programming~\cite{Lauand-2024-datafusion}.
What is new is thus the source of the contexts---interventions rather than freely chosen settings---and the shape of the resulting marginal problems, not the kind of nonclassicality.

The interventional setting also makes the role of context information explicit.
Data fusion can certify nonclassicality even when observed variables influence one another, as in a chain $A\to B\to C$, because in the causal model the edge $A\to B$ carries only the recorded value of $A$, and the intervention replaces that value by an externally chosen one.
In an experiment this has to be ensured.
If the physical link also carries anything else---the measurement context, say---the intervention does not cut it, and a classical share informed of the context can reproduce the data, as in Sec.~\ref{sec:band}.

\section{Conclusion}

Across Bell, Kochen--Specker, and causal settings the classical reading of the data is the existence of one joint distribution with the observed marginals, and quantum nonclassicality is its absence.
A single context never decides the matter; only intertwined, cyclic families of contexts can, and Farkas's lemma, as used by Garg and Mermin, turns every such failure into an inequality.
A classical model can still reproduce the data if it is allowed to know the context, which is what the cost of simulation measures.
New sources of contexts, such as interventions on causal structures, provide new instances of this obstruction rather than new kinds of nonclassicality.

\begin{acknowledgments}
During manuscript preparation, the author used Claude Code as well as Gemini to assist with literature organization,
LaTeX editing, and exact symbolic checks;
all AI-assisted output was directed and critically reviewed by the author,
 who independently verified the mathematics and takes full responsibility for the final content, and no AI tool is an author.
This research was funded in part by the Austrian
Science Fund (FWF), Grant DOI \href{https://doi.org/10.55776/PIN5424624}{10.55776/PIN5424624}.
The authors acknowledge FWF and TU Wien Bibliothek for financial support through its Open Access Funding Programme.

The author declares no conflict of interest.

No data were generated.
\end{acknowledgments}

\bibliography{svozil}

\end{document}